\documentclass[aps,prl,showpacs,amsmath,amsfonts,amssymb,twocolumn]{revtex4} 
\usepackage{psfrag}
\usepackage[dvips]{graphicx}
\usepackage{bbm}

\DeclareMathOperator{\2F1}{_2F_1}
 
\begin{document} 
 
\title{Nonanalyticities of entropy functions of finite and infinite systems} 

\author{Lapo Casetti} 
\email{lapo.casetti@unifi.it} 
\affiliation{Dipartimento di Fisica and Centro per lo Studio
delle Dinamiche Complesse (CSDC), Universit\`a di 
Firenze,\\ \mbox{and Istituto Nazionale di Fisica Nucleare (INFN), sezione di Firenze,
via G.~Sansone 1, I-50019 Sesto Fiorentino (FI), Italy}}  

\author{Michael Kastner}  
\email{michael.kastner@uni-bayreuth.de} 
\affiliation{Physikalisches Institut, Universit\"at Bayreuth, D-95440 Bayreuth,
Germany}  

\date{May 16, 2006}
 
\begin{abstract}
In contrast to the canonical ensemble where thermodynamic functions are smooth for all finite system sizes, the microcanonical entropy can show nonanalytic points also for finite systems, even if the Hamiltonian is smooth. The relation between finite and infinite system nonanalyticities is illustrated by means of a simple classical spin-like model which is exactly solvable for both, finite and infinite system sizes, showing a phase transition in the latter case. The microcanonical entropy is found to have exactly one nonanalytic point in the interior of its domain. For all finite system sizes, this point is located at the same fixed energy value $\varepsilon_{c}^{\text{finite}}$, jumping discontinuously to a different value $\varepsilon_{c}^{\text{infinite}}$ in the thermodynamic limit. Remarkably, $\varepsilon_{c}^{\text{finite}}$ equals the average {\em potential}\/ energy of the infinite system at the phase transition point. The result, supplemented with results on nonanalyticities of the microcanonical entropy for other models, indicates that care is required when trying to infer infinite system properties from finite system nonanalyticities.
\end{abstract}
 

\maketitle 

Phase transitions, like the boiling and evaporating of water at a certain temperature and pressure, are common phenomena both in everyday life and in almost any branch of physics. Loosely speaking, a phase transition brings about a sudden change of the macroscopic properties of a many-particle system while smoothly varying a parameter (the temperature or the pressure in the above example). The mathematical description of phase transitions is conventionally based on (grand)canonical thermodynamic functions, relating their loss of analyticity to the occurrence of a phase transition. Such a nonanalytic behavior in a (grand)canonical thermodynamic function can occur only in the thermodynamic limit in which the number of degrees of freedom $N$ of the system goes to infinity \cite{Griffiths}.

Different statistical ensembles, like the microcanonical or the canonical ones, lead in general to different results for statistical averages. Only in the thermodynamic limit and under suitable conditions on the interparticle interactions (namely {\em stability}\/ and {\em temperedness}\/ \cite{Gallavotti}), equivalent results are obtained in different ensembles, and one speaks of ensemble equivalence (see \cite{TouElTur:04} for an introduction or \cite{ElHaTur:00} for an extensive treatise). For a long time, interest in statistical physics concentrated almost exclusively onto the canonical and the grandcanonical ensembles, in which explicit calculations typically are easier to perform. This focus is mirrored by the fact that even the above mentioned and widely accepted definition of a phase transition makes reference to these ensembles.

For large classes of many-particle systems, e.\,g.\ in solid state physics, the thermodynamic limit is an excellent approximation, and one can profit from ensemble equivalence by performing calculations within the ensemble which appears to be the most convenient \cite{Griffiths}. However, several of the current hot topics in physics deal with systems of intermediate size, like polymers, nanosystems, or biomolecules. Although these systems are large enough for a meaningful statistical treatment, ensemble equivalence cannot be taken for granted, and it is this fact which accounts for the renewed interest in ensemble nonequivalence and finite system effects.

In the case of ensemble nonequivalence, instead of choosing the statistical ensemble according to convenience, this choice has to reflect the physical situation of interest. When a finite system is, say, energetically isolated, it has to be treated within the microcanonical ensemble, as was done for example when analyzing the experiments on sodium clusters reported in \cite{Haberland:01}. Then, in analogy to the definition of a phase transition as a nonanalyticity of a (grand)canonical thermodynamic function, it appears natural to investigate the analyticity properties of {\em microcanonical}\/ thermodynamic functions. Astonishingly, little is known about the analyticity properties of microcanonical entropy functions of finite systems.

The very recent observation that the microcanonical entropy of a finite system is not necessarily analytic, i.\,e., not necessarily infinitely many times differentiable on its entire domain, came as a surprise even to many experts in the field: nonanalytic microcanonical entropy functions of finite systems have been reported for classical \cite{KaSchne:06,DunHil1:06} as well as quantum systems \cite{BroHoHu}, where the latter case relies on a suitable but rather unconventional definition of the density of states. Note that such nonanalyticities occur for perfectly smooth Hamiltonians, so they are not introduced artificially.
% possibly: describe relation of nonanalyticities to topology changes, both classically and quantummechanically...

Having observed nonanalyticities in the microcanonical entropy of a finite system, i.\,e., in a microcanonical thermodynamic function, the similarity to the above mentioned definition of a phase transition becomes obvious. It may not be adequate to think of such finite system nonanalyticities as phase transitions in finite systems, and we will back up this statement later in this Letter. However, in the same way as nonanalyticities of (grand)canonical thermodynamic functions mark ``points of particular interest'' (i.\,e., phase transitions), the nonanalytic points of the microcanonical entropy deserve special attention and, at least in very small systems, should be measurable experimentally.

Of course not only the microcanonical entropy of a finite system can have nonanalyticities, but also its infinite system's equivalent. In contrast to the finite system ones, nonanalyticities of the infinite system entropy {\em are}\/ in general related to phase transitions, and a classification of a whole zoo of such transitions can be found in \cite{BouBa:05}.

The aim of the present Letter is to discuss the relation between nonanalyticities of the microcanonical entropy of finite and of infinite systems. This will be done mainly by discussing a spherical model with mean field-type interactions in the presence of a kinetic contribution to the energy, but the discussion will be supplemented by results from other models as well. This kinetic mean-field spherical model is one of the rare examples (the only?) which shows a proper phase transition in the microcanonical as well as in the canonical ensemble and which, in both ensembles, allows for an exact solution not only in the thermodynamic limit, but also for all finite system sizes. These properties make it an ideal laboratory for our study of analyticity properties, and to this purpose we start by presenting its microcanonical solution.

\begin{figure}[hbt]
\center
\psfrag{e}{{\normalsize $\varepsilon$}}
\psfrag{s}{{\normalsize $s$}}
\psfrag{4}{$4$}
\psfrag{8}{$8$}
\psfrag{16}{$16$}
\psfrag{inf}{$\infty$}
\psfrag{-0.5}{$-0.5$}
\psfrag{0.5}{$0.5$}
\psfrag{1.5}{$1.5$}
\psfrag{1.0}{$1.0$}
\includegraphics[width=8.3cm,height=6.6cm,keepaspectratio=true]{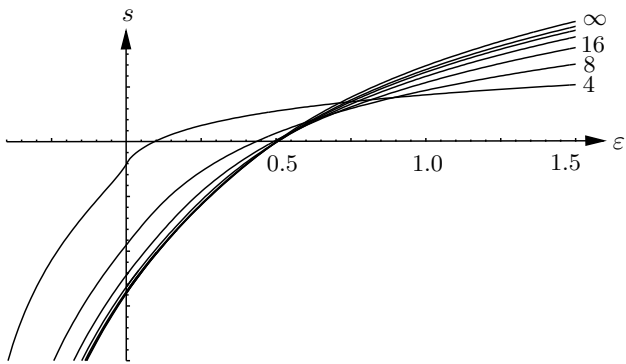}
\caption{\label{fig:entropy} \small
The graph of the microcanonical entropy $s_N$ as a function of the energy $\varepsilon$, plotted for the system sizes $N=4$, 8, 16, 32, 64, 128, and for the infinite system.}
\end{figure}

{\em Microcanonical solution of the kinetic mean-field spherical model.}--- Introduced by T.\ H.\ Berlin and M.\ Kac \cite{BerKac:52} in 1952, the spherical model of a ferromagnet is devised such as to mimic some features of the Ising model while, at the same time, being exactly solvable in the thermodynamic limit for arbitrary spatial dimensions of the lattice. We consider a mean field-like simplification of the original model where, instead of nearest-neighbor interactions, all degrees of freedom interact with each other at equal strength. The potential energy of this model is given by
\begin{equation}
V_N(\sigma)=-\frac{1}{2N}\sum_{i,j=1}^N \sigma_i \sigma_j,
\end{equation}
where the $N\geqslant2$ degrees of freedom $\sigma_i\in{\mathbbm R}$ ($i=1,...,N$) are subject to the constraint
\begin{equation}\label{eq:constraint}
\sum_{i=1}^N \sigma_i^2=N.
\end{equation}
This constraint restricts the space $\Lambda_N$ of allowed configurations $\sigma=(\sigma_1,\dotsc,\sigma_N)\in\Lambda_N$ to an ($N$--1)-sphere with radius $\sqrt{N}$. Different from the discrete Ising spin variables, the spherical model has a {\em continuous}\/ configuration space. This allows us to endow the system with a Lagrangian or Hamiltonian dynamics. Then the kinetic energy is given by the quadratic form
\begin{equation}
T_N(\dot{\sigma})=\frac{1}{2}\sum_{i=1}^N \dot{\sigma_i}^2,
\end{equation}
where the velocity vector $\dot{\sigma}=(\dot{\sigma}_1,\dotsc,\dot{\sigma}_N)$ is an element of the tangent bundle of the configuration space $\Lambda_N$. For a thus defined model, the microcanonical density of states as a function of the energy per degree of freedom $\varepsilon$ can be written as
\begin{multline}\label{eq:Omega1}
\Omega_N(\varepsilon) = a_N \int_{{\mathbbm R}^N} {\mathrm d}\sigma \int_{{\mathbbm R}^N} {\mathrm d}\dot{\sigma}\\ \times\delta\!\left(\sum_{i=1}^N \sigma_i^2-N\right)\,\delta\!\left(\sum_{i=1}^N \sigma_i \dot{\sigma}_i\right)\,\delta\!\left(T_N+V_N-N\varepsilon\right),
\end{multline}
with some normalization constant $a_N$. The first Dirac $\delta$-distribution in \eqref{eq:Omega1} accounts for the spherical constraint \eqref{eq:constraint}, the second ensures the velocity to be tangent to the configuration space, and the third $\delta$ foliates phase space into shells of constant total energy $T_N+V_N$. The form of the integral suggests the use of $N$-dimensional spherical coordinates for both, the $\sigma$ and the $\dot{\sigma}$ integrations, and an appropriate rotation of the coordinate axes renders, in each case, all but one of the angle integrations trivial. This yields
\begin{multline}\label{eq:Omega2}
\Omega_N(\varepsilon) \propto \int_0^\infty {\mathrm d}r\, r^{N-1}\,\delta\left(r^2-N\right) \int_0^\infty {\mathrm d}\dot{r}\, \dot{r}^{N-1}\\
\times\int_0^\pi {\mathrm d}\vartheta\,\sin^{N-2}\vartheta \int_0^\pi {\mathrm d}\dot{\vartheta}\,\sin^{N-2}\dot{\vartheta}\\
\times\delta\left(r\dot{r}\cos\dot{\vartheta}\right)\,\delta\left(r^2\cos^2\vartheta-\dot{r}^2+2N\varepsilon\right),
\end{multline}
where the trivial angle integrations have been executed. Three of the remaining four integrations can be performed by making use of the properties of the Dirac distribution, and the microcanonical density of states can be written in the form
\begin{multline}\label{eq:Omega_integral}
\Omega_N(\varepsilon) \propto \int_0^1 {\mathrm d}y\,y^{-1/2}\left(1-y\right)^{(N-3)/2}\\
\times\left(2\varepsilon+y\right)^{(N-3)/2} \Theta\left(2\varepsilon+y\right),
\end{multline}
where $\Theta$ denotes the Heaviside step function. This integral can be expressed in terms of Gamma functions $\Gamma$ and Gauss hypergeometric functions $\2F1$ \cite{MaObSo},
\begin{widetext}
\begin{equation}\label{eq:Omega_final}
\Omega_N(\varepsilon) \propto \left\{
\begin{array}{lll@{\quad\text{for}\;}r@{\varepsilon}l}
0 & & & & \;\leqslant-\frac{1}{2},\\
\Gamma\left(\frac{N-1}{2}\right)\Gamma\left(\frac{N}{2}\right) & \left(1+2\varepsilon\right)^{N-2} & \2F1\left(\tfrac{1}{2},\tfrac{N-1}{2},N-1;1+2\varepsilon\right) & -\frac{1}{2}<\; & \;\leqslant0,\\
\sqrt{\pi}\,\Gamma\left(N-1\right) & \left(2\varepsilon\right)^{(N-3)/2} & \2F1\left(\tfrac{1}{2},\tfrac{3-N}{2},\tfrac{N}{2};-\tfrac{1}{2\varepsilon}\right) & 0<\; & ,
\end{array}
\right.
\end{equation}
\end{widetext}
or in terms of Legendre functions. In the interior of its support $\left[-\frac{1}{2},\infty\right)$, the microcanonical density of states $\Omega_N(\varepsilon)$ has precisely one nonanalytic point at $\varepsilon_c^{\text{finite}}=0$ for all system sizes $N\geqslant2$, and the same of course holds true for the microcanonical entropy defined as
\begin{equation}
s_N(\varepsilon)=\frac{1}{N}\ln\Omega_N(\varepsilon);
\end{equation}
see Fig.~\ref{fig:entropy} for a plot.
The nonanalyticity is such that only the first $\left\lfloor\frac{N-3}{2}\right\rfloor$ derivatives of $s_N$ are continuous, where $\lfloor\cdot\rfloor$ denotes the largest integer smaller than or equal to its argument. So we see that, although a nonanalyticity is present in the microcanonical entropy for any finite system size, the strength of the nonanalyticity decreases with increasing $N$.

{\em Thermodynamic limit.}--- To compare the above described finite system behavior with that of the infinite system, the microcanonical entropy is computed in the thermodynamic limit,
\begin{equation}
s_\infty(\varepsilon)=\lim_{N\to\infty}s_N(\varepsilon)=\lim_{N\to\infty}\frac{1}{N}\ln\Omega_N(\varepsilon).
\end{equation}
Large $N$ asymptotic expansions of the Gauss hypergeometric functions occurring in \eqref{eq:Omega_final} can be found in \cite{OldeDaalhuis:03} for $\2F1\left(\cdot,\tfrac{3-N}{2},\tfrac{N}{2};\cdot\right)$ and in \cite{MaObSo} making use of the identity $\sqrt{1-2\varepsilon}\2F1\left(\tfrac{1}{2},\tfrac{N-1}{2},N-1;1+2\varepsilon\right)=\sqrt{2}\2F1\Big(\tfrac{1}{4},\tfrac{3}{4},\tfrac{N}{2};\big(\tfrac{1+2\varepsilon}{1-2\varepsilon}\big)^2\Big)$. To obtain the leading behavior only, an evaluation of the integral \eqref{eq:Omega_integral} by Laplace's method \cite{BenOrs} is sufficient, yielding, apart from a physically irrelevant additive constant,
\begin{equation}\label{eq:entropy}
s_\infty(\varepsilon)=
\begin{cases}
\ln\frac{1+2\varepsilon}{2} & \text{for $\varepsilon\leqslant\frac{1}{2}$},\\
\frac{1}{2}\ln(2\varepsilon) & \text{for $\varepsilon > \frac{1}{2}$};
\end{cases}
\end{equation}
see Fig.~\ref{fig:entropy} for a plot. A nonanalyticity in $s_\infty(\varepsilon)$ occurs at $\varepsilon_c^{\text{infinite}}=\frac{1}{2}$, and we obtain $\varepsilon_c^{\text{finite}}\neq\varepsilon_c^{\text{infinite}}$. Only the first derivative of $s_\infty$ is continuous, so a continuous phase transition is found to take place at the critical energy $\varepsilon_c^{\text{infinite}}$.

{\em Nonanalyticities and the configurational entropy.}--- In a calculation similar to that leading to Eqs.~\eqref{eq:Omega_final} and \eqref{eq:entropy}, the microcanonical average potential energy $\langle v\rangle$ can be determined as a function of the energy $\varepsilon$, and we find a critical value of the average potential energy $\langle v\rangle_c^{\text{infinite}}\equiv \langle v\rangle(\varepsilon_c^{\text{infinite}})=0$. It may seem a mere chance that this value of the critical average potential energy coincides with the energy (not average potential energy!) of the {\em finite}\/ system nonanalytic point, i.\,e.,
\begin{equation}
\langle v\rangle_c^{\text{infinite}}=\varepsilon_c^{\text{finite}},
\end{equation}
but it is not, and in fact it is a remarkable observation. The origin of this coincidence can be understood from the {\em configurational}\/ entropy $s_{\text{conf}}$ of the kinetic mean-field spherical model as a function of the potential energy $v$. This quantity, by definition, is identical to the entropy of the mean-field spherical model without a kinetic energy term as derived in \cite{KaSchne:06}. The crucial feature of $s_{\text{conf}}(v)$ is that it has compact support, and there exists a largest value $v_{\text{max}}=0$ of the potential energy accessible for the system. This value $v_{\text{max}}$ then is responsible for {\em both}, the nonanalyticity at $\varepsilon_c^{\text{finite}}=0$ in the finite system and at $\varepsilon_c^{\text{infinite}}=\frac{1}{2}$ in the infinite system: In the finite system, when increasing the energy across the value $\varepsilon=v_{\text{max}}$, suddenly no new configurations $\sigma$ become accessible, leading to the nonanalyticity in the finite system entropy $s_N(\varepsilon)$ at $\varepsilon_c^{\text{finite}}=0$. In the infinite system, in contrast, the nonanalyticity in $s_\infty(\varepsilon)$ is a consequence of the {\em average}\/ potential energy $\langle v\rangle$ reaching the value $v_{\text{max}}$ from which on {\em on average}\/ no new configurations become accessible anymore. As reasoned above, this happens at an energy $\varepsilon_c^{\text{infinite}}=\frac{1}{2}$.

{\em Phase transitions in finite and infinite systems?}--- We remarked in the introduction of this Letter that it may not be adequate to think of the nonanalyticities observed in the microcanonical entropy $s_N$ as finite system phase transitions. This statement by now should have become clear from the results on the kinetic mean-field spherical model: The nonanalytic points of the microcanonical entropy of finite systems become {\em weaker}\/ with increasing system size $N$ (in the sense that the first discontinuity arises in a derivative of order $\frac{N}{2}$). Moreover, their {\em locus}, i.\,e., the energy $\varepsilon_c^{\text{finite}}$, does not even converge towards the critical energy $\varepsilon_c^{\text{infinite}}$ of the infinite system. This, in our opinion, is not what one would expect from a reasonable generalization of the concept of a phase transition to finite systems.

{\em Nonanalytic points of the microcanonical entropy in other models.}--- To get a more complete---and at the same time more complicated---picture, we want to supplement our findings with results on the number and the quality of nonanalytic points of the microcanonical entropy in other models. To do so, it is helpful to notice the connection between nonanalyticities of the microcanonical entropy of finite systems and topology changes within the family $\left\{\Sigma_v\right\}_{v\in{\mathbbm R}}$ of constant potential energy submanifolds
\begin{equation}
\Sigma_v=\left\{\sigma\in\Lambda_N\,\big|\,V_N(\sigma)=vN\right\}.
\end{equation}
In recent studies on the relation between phase transitions and topology changes in configuration space (see \cite{Kastner:06} for an up-to-date list of references), the topology changes within the family $\left\{\Sigma_v\right\}_{v\in{\mathbbm R}}$ have been computed for several statistical mechanical models. For different models like the mean-field $\varphi^4$ model \cite{BaroniPhD}, the mean-field $XY$ model \cite{CaPeCo:03}, the mean-field $k$-trigonometric model \cite{Angelani_etal:03}, and the spherical model with nearest-neighbor interactions \cite{RiRiSta:05}, a large number of topology changes was found, increasing unboundedly with the system size $N$ and becoming dense on some interval of potential energies $v$ in the thermodynamic limit. Each of these topology changes gives rise to a nonanalyticity in the finite system microcanonical entropy. Resorting to Morse theory, it can be argued on general grounds that, again, the strength of the nonanalyticities in $s_N$ decreases with increasing $N$ such that a first discontinuity shows up in a derivative of order $\frac{N}{2}$ \cite{FranzosiPhD}. These observations make the idea of defining a finite system analogue of a phase transition via the nonanalyticities of the microcanonical entropy---an option which we have already dismissed for other reasons---appear even more inadequate: transition points, increasing unboundedly in number with the system size, do not seem to match the physical intuition of what a phase transition in a finite system should be like.

{\em Summary.}--- We have obtained an exact expression for the microcanonical entropy of the kinetic mean-field spherical model for both finite and infinite system sizes. The microcanonical entropy as a function of the energy $\varepsilon$ is found to have exactly one nonanalytic point in the interior of its domain. For all finite system sizes $N$, this point is located at the same fixed energy value $\varepsilon_{c}^{\text{finite}}=0$, jumping discontinuously to a different value $\varepsilon_{c}^{\text{infinite}}=\frac{1}{2}\neq \varepsilon_{c}^{\text{finite}}$ in the thermodynamic limit of infinite system size. Although at different values of the energy, the nonanalytic points in both, the finite system and the infinite system cases, are argued to be related to the same nonanalyticity in the configurational entropy $s_{\text{conf}}(v)$ of the system at a potential energy $v=0$.

Two important conclusions can be drawn from these peculiar findings: (i) Finite system and infinite system nonanalyticities of the microcanonical entropy are not unrelated, but their relation is not a straightforward one. Care is required when trying to infer infinite system properties from finite system nonanalyticities. (ii) A generalization of the definition of a phase transition via the nonanalyticities of the microcanonical entropy is too na\"{\i}ve an approach: the number of such nonanalyticities may incease unboundedly with the system size, whereas their strength may decrease.

We would like to thank Oliver Schnetz for suggesting the change of coordinates employed to arrive at Eq.~\eqref{eq:Omega2}. L.~C.\ and M.~K.\ acknowledge financial support from the PRIN05-MIUR project {\em Dynamics and thermodynamics of systems with long-range interactions}. M.\,K.\ acknowledges financial support by the Deutsche Forschungsgemeinschaft (grant KA2272/2).

\bibliographystyle{h-physrev}
\bibliography{SphericalKinetic}

\end{document}